\documentclass[a4paper,11pt]{article}
\usepackage{nfraconf,graphicx,cite,psfig}

\begin{document}
\baselineskip=13pt

\title{Studying High Redshift Star Forming Galaxies at \\
Centimeter and Millimeter Wavelengths} 

\author{C.~L.Carilli$^{1,2}$, 
K.~M. Menten$^2$,  M.~S. Yun$^{1}$, F. Bertoldi$^{2}$, 
F. Owen$^{1}$, A. Dey$^{3}$}

\maketitle

~$^{1}$National Radio Astronomy Observatory, Socorro, NM, USA,
ccarilli@nrao.edu 

~$^2$Max-Planck-Instit\"ut f\"ur Radioastronomie, Bonn, Germany

~$^3$National Optical Astronomy Observatory, Tucson, AZ, USA

\vskip 0.2in

\abstract{
We discuss various aspects of centimeter and millimeter
wavelength continuum and line observations
of high redshift star forming galaxies.  Perhaps the most
important lesson is that sensitive observations at submm through cm
wavelengths reveal a population of active star forming galaxies at
high redshift which are unseen in deep optical surveys due to dust
obscuration.  Current
models suggest that this population represents the formation of the
spheroidal components of galaxies at $z$ between 2 and 5, 
constituting about half of the total amount of cosmic star formation
from the big bang to the present. 
High resolution imaging at cm wavelengths provides
sub-arcsecond astrometry, and can be used to search for gravitational
lensing and/or for the presence of an 
AGN. Radio continuum observations provide unique
information on the magnetic fields in early galaxies, and give a gross
indication of the star formation rate, while the radio-to-submm spectral
index provides a rough indication of source redshift. Low
$J$ transitions of CO are redshifted into the cm bands for $z > 2$,
allowing for sensitive searches for CO emission over large
volumes at high redshift.  We present recent results from the
Very Large Array (VLA), and from the new 230 GHz MPIfR bolometer array
at the IRAM 30m telescope. A wide field survey with the bolometer
array indicates a cut-off in the source distribution function 
at FIR luminosities $>$  $3 \times 10^{12}$ L$_\odot$.
Lastly,  we summarize the scientific promise of  the New VLA. 
}

\vskip 0.15in

To appear in {\sl Scientific Imperatives at centimeter and meter
Wavelengths}, ed. M.P van Haarlem and J.M. van der Hulst, 
(Dwingeloo: NFRA).

\section{Introduction}

The sharp rise of observed flux density, $S_\nu$, with increasing
frequency, $\nu$, 
in the Rayleigh-Jeans portion of the grey-body spectrum
for thermal dust emission from star forming galaxies ($S_\nu \propto
\nu^{3.25}$), leads to a dramatic negative $K$-correction for observed flux
density with increasing redshift. Submm surveys thereby provide a
uniquely {\it distance independent} sample of objects in the universe,
meaning the observed submm flux density of an object of given intrinsic
luminosity is roughly constant, or even increases substantially for certain
cosmologies,  for $z$ between 1 and
5 for a fixed observing frequency (Blain \& Longair 1993).

Initial studies of optical galaxies in the Hubble Deep Field suggested
a sharp peak in the cosmic star formation density at around $z$ = 1.5
(Madau et al. 1996). However, recent observations of faint submm
source counts 
argue that the cosmic star formation rate may be roughly constant
between $z$ = 1 to 5, implying that optical studies may be missing a
substantial amount of star formation occurring in dust-obscured
starbursts at high $z$ (Hughes et al. 1998, Eales et al. 1999, Blain et
al. 1999a).  This notion appears to have been confirmed through new
optical observations of high $z$ galaxies by including larger
dust-extinction corrections (Steidel et al. 1999, Calzetti \& Heckman
1999). 

Tan, Silk, \&\ Blanard (1999) have presented a model of the star
formation history of the universe which may explain the faint source
counts from the optical through the radio regimes. They hypothesize
that star formation occurs in two phases. At high redshift ($2 < z <
5$), very active star formation, with rates $\ge$ 100 M$_\odot$
year$^{-1}$, occurs in merger events leading to the
formation of the spheroidal components of galaxies (spiral bulges and
ellipticals; Barger, Cowie, \& Sanders 1999). At lower redshift ($ z<
2$), the stellar disks of 
galaxies are formed over longer periods at lower star formation rates.
The high $z$ starbursts are dusty systems and are revealed primarily
as constituents of the faint submm, IR, and radio source populations.
In contrast, the faint optical source counts are dominated by the
lower $z$ disk-forming systems.  Tan et al. state that the early
spheriods may act as ``seeds'' for disk formation as gas falls in, and
that the present-day stellar mass in bulges and halos is comparable to
that in disks, i.e. that about half of the total amount
of the star formation in the universe has occurred in these dusty, high
$z$ starbursts. Moreover, this population of high $z$ star-forming galaxies
may provide a significant contribution to cosmic 
metal production (Fall \& Pei 1999), and to the 
re-ionization of the universe at $z > 3$ (Baltz, Gnedin, \& Silk
1998). 

Given the tight radio-to-far infrared (FIR)
correlation in star-forming galaxies
(Condon 1992), the mJy submm source population can
be equated with, for the most part, the $\mu$Jy radio source
population. Observations at cm wavelengths can provide important
information on high $z$ star-forming galaxies, complementary to mm,
submm, and near IR observations:
\begin{itemize}

\item{Radio interferometry offers sub-arcsecond astrometry, which is
critical for faint source identification in confused optical and near
IR fields (Blain 1999, Smail et al. 1999).}

\item{The submm-to-cm spectral
index provides a rough indication of source redshift (Carilli \& Yun
1999, Blain 1999). The redshift distribution of the faint submm source
population is a key parameter for constraining models of the star
formation history of the universe using submm source counts, 
and presently a point of serious debate 
(Blain et al. 1999b). }

\item{High resolution radio imaging provides information on a number
of issues, including: (i) the existence of an active nucleus, (ii) the
distribution of star formation in early galaxies, and (iii) the possibility
of gravitational lensing.}

\item{Radio continuum emission contains unique information on the
strength and structure of magnetic fields in early galaxies (Beck et
al. 1996).} 

\item{Radio continuum emission provides an indication of the star
formation rate, via the tight radio-to-FIR correlation for star
forming galaxies (Condon 1992).}

\item{CO emission involving 
low rotational quantum numbers
($J$) are redshifted to cm wavelengths for $z \ge 2$.}

\end{itemize}

In this contribution we discuss various aspects of continuum and line
observations of high redshift star forming galaxies at cm wavelengths.
We present recent results from the Very Large Array, and discuss the
future capabilities after the  VLA expansion (the `New VLA'). 
We assume $H_0$ = 75 km s$^{-1}$ Mpc$^{-1}$ and $q_0$ = 0.5.

\section{The Radio-to-FIR Spectral Index as a Redshift Indicator}

The most difficult aspect of studying star forming galaxies at high
redshift is determining the redshift. 
Redshifts are critical for constraining models of the
star formation history of the universe using faint submm source
counts. However, the study of the
faint submm source population shows that the more active star forming
galaxies at high redshift are highly dust-obscured, 
and hence challenging to study even with the largest optical
telescopes, with typical magnitudes of I $>$ 25, and  
colors I -- K $\ge$ 4 (Smail et al. 1999, Cimatti et al. 1998). 

We have recently considered the possibility of using the radio-to-FIR
spectral index as a gross indicator of redshift for star forming
galaxies (Carilli \& Yun 1999). Blain (1999) has subsequently re-cast
this relationship into a format involving the logarithm of flux density
ratio. The spectral index is related
to the logarithm of the flux density ratio simply by a factor 2.4 =
log[${350}\over{1.4}$].  The technique is
based on the very tight correlation found between the radio and far IR
emission from star forming galaxies, and the effect is demonstrated in
Figure 1.  The 1.4 GHz to 350 GHz spectral index will change from
about 0 at low redshift, to about 1 at $z$ $>$ 3 for a star forming galaxy
with an SED similar to M82.  This is not a subtle effect -- a unit
change in the spectral index corresponds to a change in
flux density ratio by a factor 250.  
We emphasize that there is significant scatter in
the observed value for galaxies at a given redshift, such that this
technique should be used simply as a redshift `indicator' (i.e. low $z$
versus high $z$), analogous to the optical `drop-out' technique, and not
as a redshift `estimator'.  Still, even a gross indication of the
source redshift distribution is an important model constraint (Smail
et al. 1999).

\begin{center}
\begin{figure}
\hskip 1in
\psfig{figure=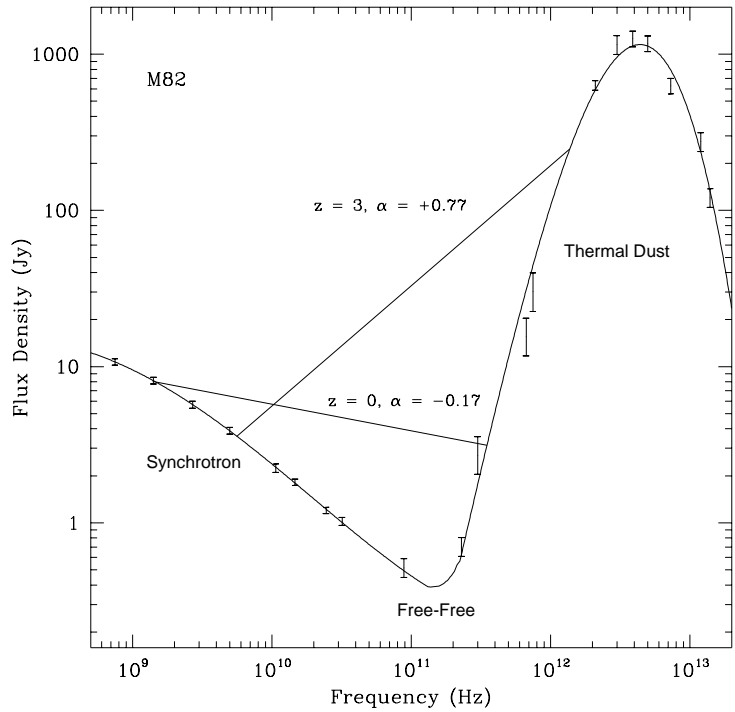,width=4in}
\hspace*{0.8in}
\psfig{figure=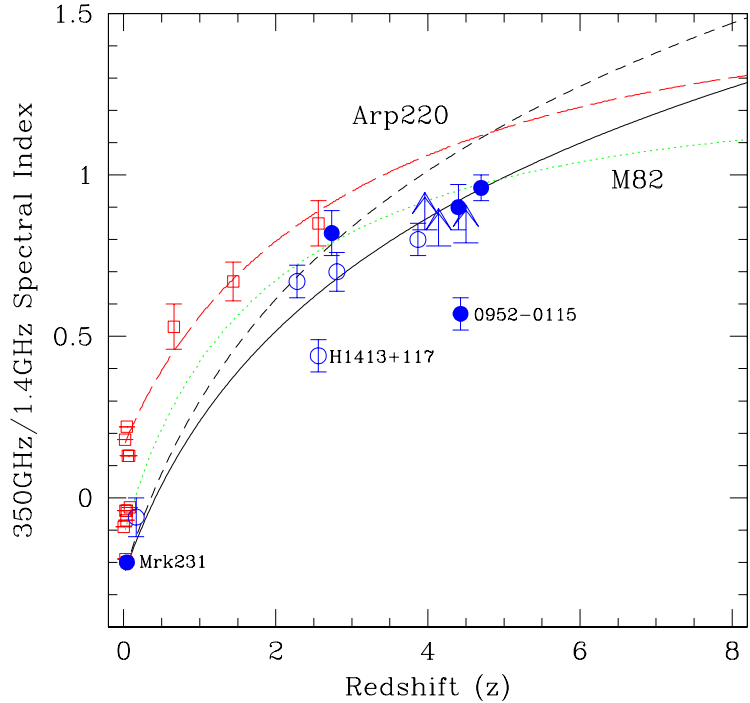,width=4.3in}
\vspace*{-0.3in}                                              
    \caption{\small The upper panel shows the
radio-thru-infrared SED of M82. The two straight lines 
show the change in  observed spectral index between 1.4 GHz and 350 GHz 
for a source at $z$ = 0 and $z$ = 3. The lower panel shows the behavior 
of the observed  1.4 GHz-to-350 GHz spectral index as a function of
source  redshift for  star forming
galaxies. The models are semi-analytic, based on equations in
Condon (1992), and empirical models based on observations of 
Arp 220 and M82. The data points show the results for 
high redshift submm sources (see Carilli \& Yun 1999, and Yun et
al. 1999  for details). Note that the position of 1413+117 and
0952--0115 on this diagram indicate the existence on a radio-loud AGN. 
}
\end{figure}
\end{center}

Carilli \& Yun (1999) and Blain (1999) discuss the origin of the
scatter in the relationship. Two possibly important effects in the
radio are free-free absorption at low frequency, and inverse Compton
losses off the microwave background at high redshift, while in the
submm the dust temperature can alter the relationship
substantially. The models in Figure 1 are for a dust temperature of
about 50 K, as is found in the starburst nucleus of M82 (G\"usten et
al. 1993). Temperatures around this value appear to be prevalent for
high $z$ starburst systems (Benford et
al. 1999). Of course, the presence of a radio loud AGN will
invalidate the relationship. Turning this around, given a source
redshift, this relationship can be used to test for the existence
of a radio loud AGN. As examples, the presence of a radio loud AGN can
be seen for the two sources 1413+117 and 0952--0115 in Figure 1.
Contamination of submm samples by radio loud AGN is thought to be
$\le$ 10$\%$ (Yun, Reddy, \& Condon in preparation). 

\section{The Hubble Deep Field}

The Hubble Deep Field (HDF) has played a dominant role in the study of
galaxy evolution over the last few years. Deep optical  imaging has
revealed thousands of galaxies in a region of only a few square minutes. 
At first glance, radio and submm observations of the HDF appear to be
disappointing -- with only a handful of firm detections at 1.4 GHz
(Richards et al. 1999), and even fewer at 350 GHz (Hughes et al. 1998).
There are many simulations in this volume which demonstrate how
the SKA will  detect most of the galaxies in the HDF with
reasonable integration times. 

However, we would like to emphasize that the existing cm and submm 
data on the HDF already have a profound consequence:
thus far there is no convincing optical identification of the brightest
submm source in the HDF (Downes et al. 1999).
Based on deep imaging at 230 GHz using the Plateau de Bure
interferometer,  and VLA
imaging at 1.4 GHz, the submm and radio source 
is situated  between an elliptical galaxy at $z$ = 1.12 and
a faint arc of emission with an unknown redshift (Downes et al. 1999, 
Figure 2). The source is 7 mJy at 350 GHz (Hughes et al. 1998),
and 15 $\mu$Jy at 1.4 GHz (Richards et al. 1999), 
implying $z$ $>$ 3 based on the relationship shown in Figure 1. 

\begin{center}
\begin{figure}
\hskip 0.8in
\psfig{figure=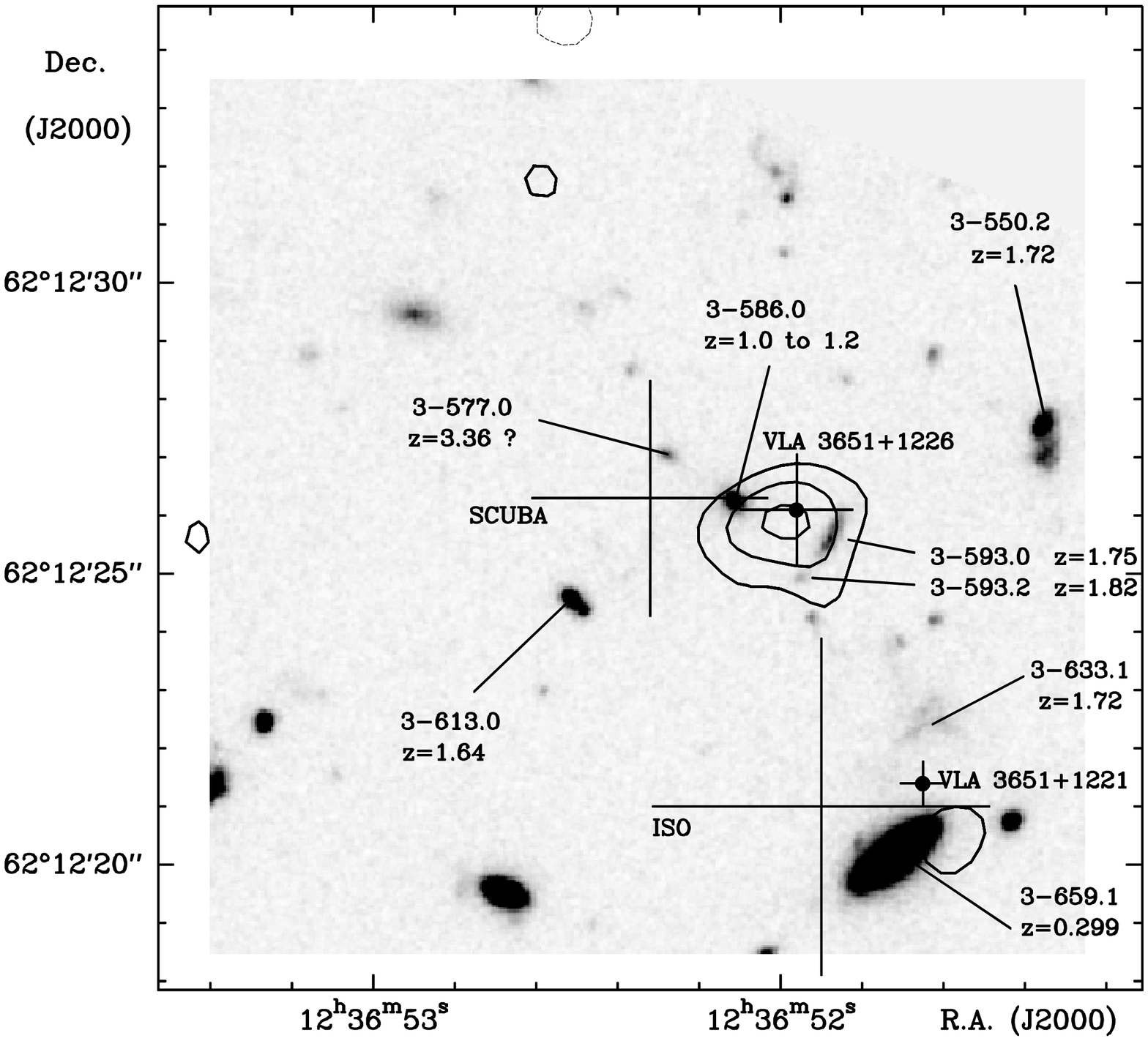,width=4in,angle=-90}
    \caption{\small The contours show the 230 GHz continuum emission
    from HDF 850.1 as observed with the IRAM interferometer
    (Downes et al. 1999). The grey scale is the HST image of the
    field. A cross marks the position of the radio continuum source. 
    The redshifts indicated are very tentative, as derived in
    Downes et al. (1999) from marginal optical broad band photometry.
    These require  confirmation. Radio and submm
    data suggest a large redshift for the 230 GHz source (z $\ge$ 3).
}
\end{figure}
\end{center}

This observation of the HDF is a dramatic demonstration that the mJy
submm and $\mu$Jy radio source counts reveal a population of sources
that are essentially unseen in the deepest optical observations. Until
the advent of the NGST, systematic study of a large sample of these
sources will be the exclusive regime of radio through infrared
observations (Blain 1999), with optical follow-up observations limited
to a few objects using long integrations on the largest optical
telescopes.  It also emphasizes the extreme difficulty and care that
must be taken when making optical identifications of faint submm
sources based on low spatial resolution data. Deep
radio imaging with sub-arcsecond astrometry is critical to any optical
follow-up program to study dust obscured, high redshift star forming
galaxies.

\section{Star Formation and Massive Black Holes at High Redshift}

Omont et al. (1996) have made the remarkable discovery that a
significant fraction of optically selected high $z$ QSOs show thermal
dust emission.  Of their sample of 16 QSOs at $z > 4$, a total of 6
sources show dust emission at 230 GHz, with implied dust masses of
10$^8$ to 10$^9$ $M_\odot$. Follow-up mm line observations of three of
these dust emitting QSOs have revealed CO emission with implied
molecular gas masses $\sim$ 10$^{11}$ $M_\odot$ (Ohta et al. 1996;
Guilloteau et al. 1997).  Omont et
al. (1996) speculate that ``$\ldots$such large amounts of dust [and
gas] imply giant starbursts at $z > 4$, at least comparable to those
found in the most hyper-luminous IRAS galaxies$\ldots$''.  However, the
evidence for active star formation in these sources remains
circumstantial, primarily based on the presence of large gas
reservoirs. It remains possible that the dust is heated by the active
galactic nucleus (AGN), rather than by a starburst.

We have used the VLA to study the radio continuum and CO(2--1)
emission from one of these systems, 1335--0417, at $z = 4.4$.  The
CO(2--1) line redshifts into the 43 GHz band of the VLA for $4.9 > z >
3.8$ (Figure 3). We have detected CO emission at 43 GHz, and radio
continuum emission at 1.4 and 5 GHz, from 1335--0417 using the VLA
(Carilli, Menten, \& Yun 1999).
We find that the ratio of the velocity-integrated line intensity for
the CO(2--1) and CO(5--4) emission  (Guilloteau et al. 1997)
from 1335--0417 is consistent with
constant brightness temperature (i.e. integrated
line flux increasing as $\nu^2$), implying optically thick emission
in both lines, for a fixed source size. Constant brightness
temperature implies that the molecular gas must be fairly warm, with a
lower limit to the rotational temperature of 30 K set by the
excitation of the CO (5--4) line, and that the line emission is
optically thick.  A lower limit to the source size can be derived
assuming a single, homogeneous, optically thick emitting region (Ohta
et al. 1996). The beam filling factor is given by: ${\Omega_{\rm
s}}\over{\Omega_{\rm b}}$ = ${{T_{\rm obs}}\over{T_{\rm ex}}} \times
(1+z)$, where $\Omega_{\rm s}$ and $\Omega_{\rm b}$ are the source and
beam solid angles, respectively, $T_{\rm obs}$ is the observed line
brightness temperature, and the line excitation temperature, $T_{\rm
ex}$, is assumed to be equal to the intrinsic (rest frame) source
brightness temperature. Note that for a resolved source, the observed
brightness temperature obeys: $T_{\rm obs}$ = ${T_{\rm
ex}}\over{(1+z)}$. We find that the minimum source diameter for
1335--0417 is $0{\rlap.}{''}23\times({{T_{\rm ex}}\over{50
K}})^{-{1\over2}}$. 

\begin{center}
\begin{figure}
\vskip -1.8in
\psfig{figure=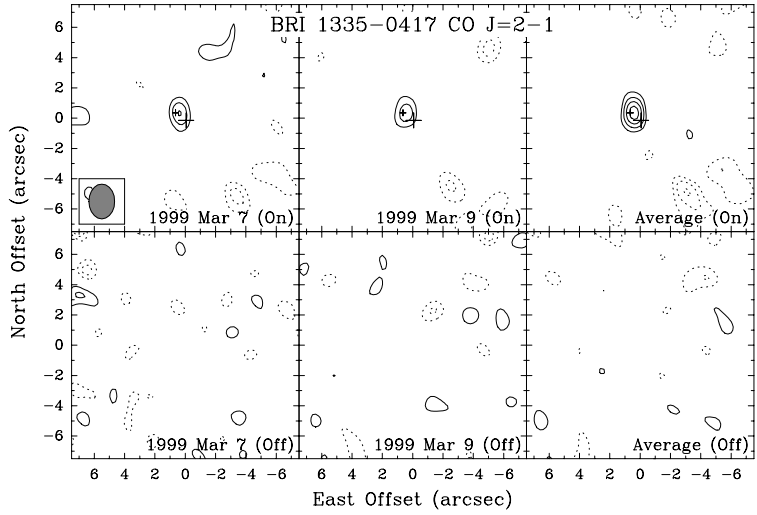,width=5in}
\vspace*{-4.2in}
\hskip -0.7in
\psfig{figure=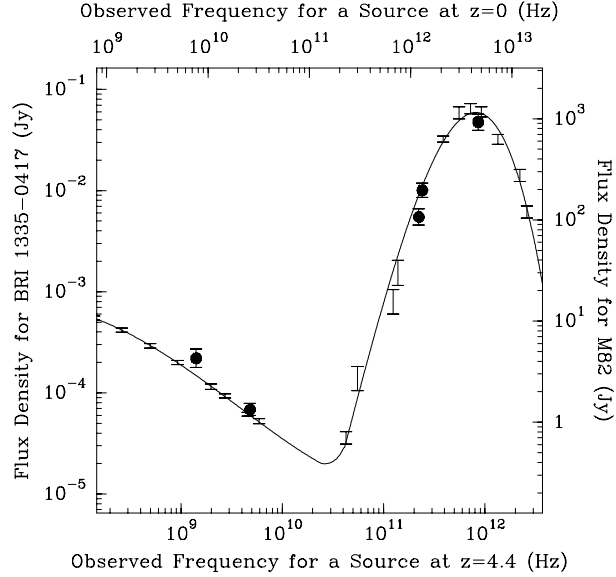,width=7in}
\vspace*{-5.3in}                                              
    \caption{\small 
The upper panel shows VLA contour maps of  1335$-$0417 taken at a
frequency centered on (upper panels), and off (lower panels) the
frequency of the redshifted ($z = 4.4074$) CO (2--1) line emission
near 43~GHz.  The leftmost and middle panels show the maps produced
from data taken on 1999 March 7 and 9, respectively, while the
rightmost map was produced from the average of these individual
datasets.  The contour levels represent $-$4, $-$3, $-$2, 2, 3, 4, and
5 times the $1\sigma$ rms noise, which is 0.19 mJy beam$^{-1}$ for the
``single day'' maps and 0.13 mJy beam$^{-1}$ for the map made from the
averaged data.  The maps have been restored with a beam of size
$2{\rlap.}{''}3\times1{\rlap.}{''}7$
(FWHM) elongated N-S, which is indicated by the
grey ellipse in the leftmost ``on'' map.  The small cross marks the
position of the 1.35 mm dust continuum source determined by Guilloteau
et al. 1997, with the size of the cross indicating the $\pm
0{\rlap.}{''}2$ uncertainty of the mm position.  The larger cross
marks the optical position given by Storrie-Lombardi et al.  1996
without errors quoted. 
In the lower panel, 
the solid line shows the radio through infrared
spectral energy distribution (SED) of M82 derived via polynomial fits
to the observed data points, which are taken from the NASA
Extragalactic Data  Base (NED) and are shown as vertical error bars. 
The solid dots (plus error bars) show the data for the
$z = 4.4074$ QSO  1335$-$0417.  For comparative purposes, the data for
 1335$-$0417 have been
normalized to the SED of M82 using the mean of the normalization value
at rest frame frequencies of 26 GHz and 1300 GHz.
The left and right hand flux density scales are those appropriate for
 1335$-$0417 and M82, respectively. 
Note that the intrinsic luminosity at 1.4 GHz of  1335$-$0417 is a
factor 1400 times larger than that of M82.
The lower abscissa is chosen to represent the observed frequency scale for
the   1335$-$0417 data points  (i.e. for a source at $z$ = 4.4),
whereas the upper abscissa is appropriate
for the M82 data points (i.e. for a source at $z$ = 0).  
}
\end{figure}
\end{center}

The critical question concerning  1335$-$0417 is whether the dust
and radio continuum emission are powered by the AGN, or star
formation, or both.  Figure 3 shows the integrated spectral energy
distribution for  1335$-$0417 for rest frame frequencies from 7.6
GHz up to 4600 GHz, normalized to the spectrum of the nuclear
starburst galaxy M82. For demonstrative purposes we have quantified
the spectrum of M82 using two accurate polynomial fits: one to the
synchrotron emission which dominates at frequencies below 100 GHz, and
a second to the thermal dust emission which dominates above 100
GHz. All the data points for  1335$-$0417 fall within 2$\sigma$ of
the M82 curve.  The agreement between the SED of  1335$-$0417 with
that of
M82 argues in favor of star formation playing the dominant role in
heating the dust and powering the radio continuum emission from
 1335$-$0417. It is important to note that, while the shapes of 
the SEDs are similar, the luminosity of  1335$-$0417 is
1.2$\times10^{32}$ ergs s$^{-1}$ Hz$^{-1}$ at 1.47 GHz in the rest
frame of the source, assuming a radio spectral index of $-0.8$, while
that of M82 is a factor 1400 lower (assuming a distance of 3 Mpc to
M82). If the emission is powered by star formation, a very rough
estimate of the massive star formation rate in  1335$-$0417 can be
derived from the radio and submm continuum emission using equations 1
and 2 in Carilli \& Yun (1999). Both values are consistent with a
massive  star
formation rate of 2300$\pm$600 $M_\odot$ yr$^{-1}$. We emphasize that
this is at best an order-of-magnitude estimate.

Overall, the cm and mm continuum observations, as well as the CO line
data, argue in favor of a very massive starburst concurrent with the
AGN activity in  1335$-$0417 at $z = 4.4074$.  A similar conclusion
has been reached concerning a number of other dust emitting QSOs from
the Omont et al. sample by Yun et al. (1999).  The gas depletion
timescale is short, $\le$ 10$^9$ years, in which time a significant
fraction of the stars in the host galaxy of the AGN may be formed 
(Benford et al. 1999).
 
\section{Bright Faint Source Survey}

A major new instrument in the study of faint mm sources  is the MPIfR
230 GHz, 37 element bolometer camera on the IRAM 30m telescope
(Kreysa et al. 1999). The sensitivity of this instrument, coupled with 
the better atmosphere at 230 GHz relative to 350 GHz, and the large
aperture of the 30m telescope makes this instrument competitive with
SCUBA on the JCMT at 350 GHz (Holland et al. 1999), 
even including the sharply rising spectrum of thermal sources. 
We had a long observing session using this camera on the 30m during
February and March, 1999. The initial results indicate that an rms
sensitivity of 0.4 mJy can be reached in 5 hours under
good conditions for on-off observations of point sources. 

One of the initial observing programs was wide field imaging of the
field around the $z$ = 0.27 cluster A2125. We chose this field because
very deep radio imaging has been performed at the VLA, to a
sensitivity level of 6$\mu$Jy at 1.4 GHz (Owen \& Morrison, in prep.).  
The existence of a cluster may help to enhance the faint source counts
through gravitational lensing (Blain et al. 1999), at least 
within a few arcmin of the cluster center.  It was decided to
image a relatively large field to a moderate sensitivity level, in
order to study the high-end of the source luminosity function (the
`Bright-Faint' sources). Also, it can be shown that as long as the
index for the source luminosity function is less than  2, 
this type of medium
sensitivity, wide-field mapping procedure is the most efficient in
terms of maximizing the number of sources detected with a given total
observing time. 
Detecting a large number of sources over a wide field will allow us 
to study source clustering properties on scales from 100 kpc to
2 Mpc, in order to test the current idea
that the faint submm sources may be the precursors of present day 
elliptical galaxies (Smail et al. 1999, Tan et al. 1999).

Smail et al. (1999) present a cumulative source 
distribution that is consistent with a power-law of the form:
N($>S$) = 1800 $\times (S_{230})^{-1.1}$ deg$^{-2}$, where $S_{230}$
is the flux density in mJy at 230 GHz (Smail et al. 1999), assuming a
submm spectral index of +3.25, leading to a factor 4 lower flux
density at 230 GHz relative to 350 GHz.  Barger,  Cowie, \&
Richards (1999) obtain a somewhat steeper distribution, with
a possible cut-off at 
$S_{230}$ $\ge$ 2 mJy, although the statistics 
at high flux levels remain limited. 
To date, we have surveyed a total area of about 100 armin$^2$ to an rms
sensitivity of $\le$ 1 mJy at 230 GHz with the MPI bolometer array at the
IRAM 30m. Using the Smail et al. (1999) source distribution, we would 
expect to detect about 25 sources  above
2 mJy, and 14 sources above 3.5 mJy.
We find about 10 sources between 2 and 3.5 mJy,
and no sources above 3.5 mJy. This is consistent with the 
possible cut-off at high luminosities suggested by Barger et
al. (1999). Note that 2 mJy at 230 GHz corresponds to 
an FIR luminosity of about $3 \times 10^{12}$ L$_\odot$, or a massive star
formation rate of about 470 M$_\odot$ year$^{-1}$ (Cowie \& Barger
1999, Carilli \& Yun 1999).  

Figure 4 shows
one of the sources we have detected in the A2125 field. The source is
3 mJy at 230 GHz. It is coincident with a 70 $\mu$Jy radio source.
Nothing is seen at the position in the optical to R $>$ 24.5.  A deep
K band image taken with the Keck telescope reveals a source at the
expected position with K = 21. Based on the radio-to-submm spectral
index, this source is likely to be at $z > 2$.
Note that within the 12$''$ (FWHM) beam of the 30m telescope
there are three faint near IR sources. The radio
continuum emission provides an unambiguous identification of the 
230 GHz source. Again, these observations are meant to illustrate the
importance of deep radio imaging with sub-arcsecond resolution for
accurate source identification, and the difficulty of follow-up
optical identification for these very faint, red sources (Cimatti et
al. 1998).

\begin{figure}
\hskip 1.6in
\psfig{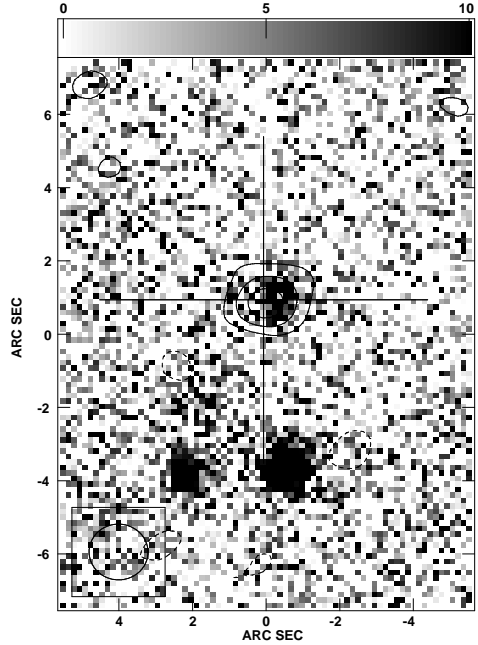}
    \caption{\small 
The contours show the 1.4 GHz emission from a 70$\mu$Jy source
in the A2125 field, with levels -36, -18, 18, 36, 54, and 72
$\mu$Jy/beam. The grey scale is a K band  image of the field from
the Keck telescope. The cross marks the position of a 3 mJy source at
230 GHz, and the cross size is the FWHM of the IRAM 30m primary beam.
}
\end{figure}

\section{Near Future: The New VLA}

The science case for studying high redshift star forming
galaxies with the Atacama Large Millimeter Array (ALMA),
is well documented (see MMA
home page on http://www.nrao.edu), while that for the SKA
is spelled-out in detail in this volume. In this section we discuss
the nearer-term possibilities that will arise with the expanded VLA
within the next 5 years. A good discussion of the synergy between the
submm, mm, and cm observations with future instrumentation can be
found in Blain (1999). 

The principal aspect of the VLA expansion is a wideband 
(4 $\times$ 2 GHz $\times$ 2
polarizations) correlator with greatly increased spectral
capabilities. The expansion also includes: (i) upgrades to existing
receivers to decrease system noise and to provide complete frequency
coverage from 1.0 to 50 GHz, (ii) improvements to antenna pointing and
efficiency, and, on a longer timescale, (iii) additional antennas
extending the maximum baseline by a factor of 10 to 300 km (see
VLA upgrade homepage on http://www.nrao.edu).  Overall, the VLA expansion
represents an improvement by an order of magnitude, or more, for all
aspects of observing with the VLA. The New VLA
propels cm continuum observation into the realm of sub-$\mu$Jy
sensitivity, while the new spectral line capabilities 
allow for large volume searches for high redshift CO
emission.

\begin{figure}
\hskip 1in
\psfig{figure=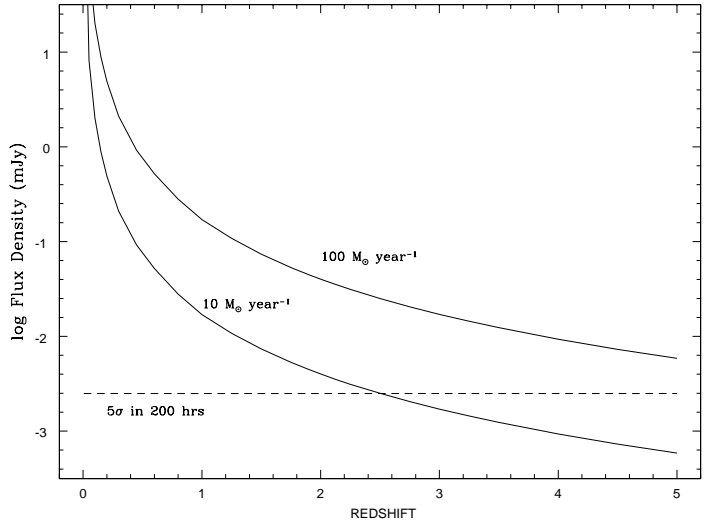,width=4in,angle=-90}
    \caption{\small 
The expected radio continuum flux density at 1.4 GHz of star forming
galaxies as a function of redshift. The flux densities were derived
using the relationships in Condon (1992). The dashed line shows the
5$\sigma$ sensitivity of the New VLA in 200 hrs.
}
\end{figure}

Figure 5 shows the expected sensitivity of the New VLA at 1.4 GHz
for a long, but not unreasonable integration time (200 hr), as
compared with the expected flux density for synchrotron emission from
typical star forming galaxies.  The New VLA will be able to
detect synchrotron emission from massive star forming galaxies (100
M$_\odot$ yr$^{-1}$) to $z \sim 5$, corresponding to sources with 350
GHz flux densities of 2 mJy, and lower star formation rate galaxies
(10 M$_\odot$ yr$^{-1}$) to $z = 2.5$. The relationship between massive
star
formation rate and radio continuum luminosity is taken from Condon
(1992).  Note that on this scale a typical ultra-luminous infrared
galaxy (ULIRG), like Arp 220, has a massive
star formation rate $\approx$ 100 M$_\odot$ yr$^{-1}$.

\begin{figure}
\psfig{figure=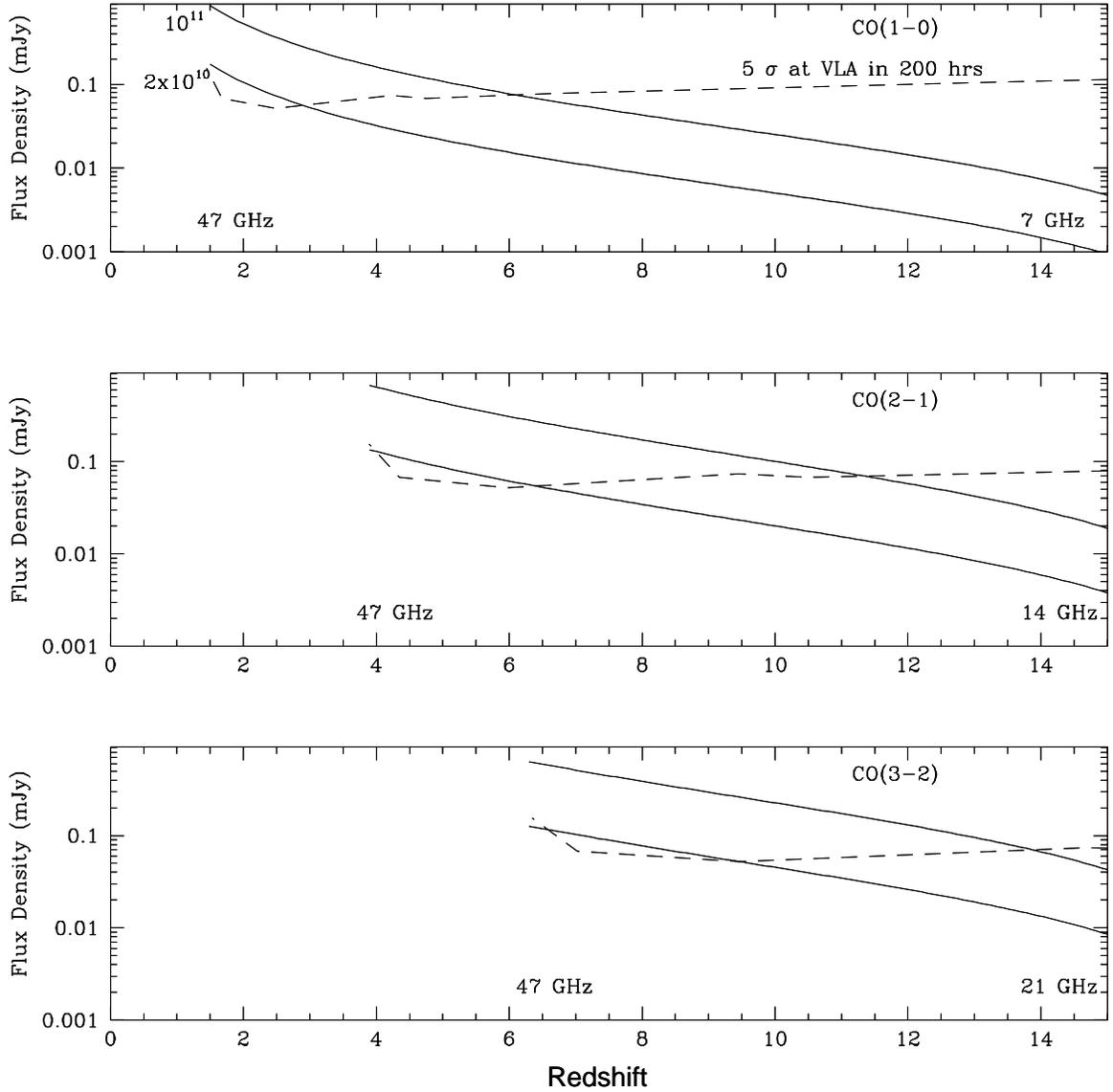,width=6.5in}
    \caption{\small 
The expected CO emission as a function of redshift for three CO
transitions. The upper solid line in each panel is for a molecular 
gas mass of 10$^{11}$
M$_\odot$, and the lower solid line is for 2$\times10^{10}$ M$_\odot$.
The flux densities as a function of mass and redshift
were derived as described
in section 6.  In terms
of CO luminosity, the ULIRGs Arp 220 and Mrk 231 would both have gas
masses of about 2$\times 10^{10}$ M$_\odot$ on the scale in Figure 6,
corresponding to the lower solid line in each panel.
The dashed line shows the New VLA sensitivity (5$\sigma$ in
200hrs at 400 km s$^{-1}$ resolution).  
The observing frequency at each end of the curve for
a given transition and redshift is
listed along the bottom of each panel. 
}
\end{figure}

A second important use of the New VLA will be to image low-$J$
CO emission from high redshift galaxies. Figure 6 shows the
sensitivity and redshift range covered by the New VLA for the
CO(1--0), CO(2--1), and CO(3--2) transitions. We have assumed constant
brightness temperature for the lines. We use the expression relating
gas mass to CO luminosity from Solomon, Radford, \& Downes (1992),
with the added factor of line `suppression' due to the microwave
background radiation.\footnote{The line brightness temperature is
measured only relative to the background, such that the observed line
flux density, $S_{\rm obs}$, is decreased relative to the `true' line
flux, $S_{\rm true}$ (i.e. in the absence of the microwave background
radiation), by the factor: $S_{\rm obs} = S_{\rm true} \times (1 -
{{T_{\rm bg}}\over{T_{\rm ex}}})$, where $T_{\rm bg} = 2.73 \times (1
+ z)$. In other words, the background radiation is `blocked' by the
source.} We adopt a gas temperature of 50 K (Benford et al. 1999). Note
that the Solomon et al. expression is derived assuming a Galactic gas
mass-to-CO luminosity conversion factor. Current observations 
suggest that this conversion factor overestimates the gas mass for extreme
starburst systems by at least a factor 3
(Carilli et al. 1999, Solomon and Downes 1998). In terms
of CO luminosity, the ULIRGs Arp 220 and Mrk 231 would both have gas
masses of about 2$\times 10^{10}$ M$_\odot$ on the scale in Figure 6.

The New VLA will be
able to detect CO emission from ULIRGs, such as Arp 220, to large
redshifts. Moreover, the New VLA will have enough sensitivity and
resolution to produce spatially resolved images of the 
CO emission from these high redshift systems. 
The New VLA will have a 1$\sigma$ brightness temperature
sensitivity in 200 hrs in the B array at  43 GHz 
of 0.25 K at 200 mas resolution for a
line with FHWM = 400 km s$^{-1}$, adequate to image the CO
emission from the $z$ = 4.4 QSO 1335--0417 
(see section 4). We should also point out that, due to
correlator limitations, the current CO observations of 1335--0417 with
the VLA were done in continuum mode, using just two channels, an
'on-line' and an 'off-line' channel each with the maximum bandwidth of
50 MHz, or 350 km s$^{-1}$.  The upgraded VLA correlator will provide
adequate bandwidth and spectral channels to obtain a proper spectrum
of the emission from 1335--0417.

Notice that the expected flux densities  for the three 
CO transitions are roughly the same at any  given frequency.
For instance, at 30 GHz the three CO transitions,
CO(1--0), (2--1), and (3--2), are seen at $z$ = 2.8, 6.7, and 10.5,
respectively, and have expected flux densities of 27$\mu$Jy, 26 $\mu$Jy, and
20$\mu$Jy, respectively  for 10$^{10}$ M$_\odot$ of gas. This is due to the
(reasonable) assumption of constant line brightness temperature. This leads
to the interesting capability that any survey for high $z$ CO emission
in these frequency ranges automatically covers a number of redshift
ranges at a fairly uniform mass sensitivity limit. As an example, 
with the New
VLA one can envision a search for CO emission using a single frequency
setting covering the range of 26 to 34 GHz.
The primary beam has FWHM =1.4$'$, and the 5$\sigma$ sensitivity at 400 km
s$^{-1}$ resolution in 200 hrs is about 70 $\mu$Jy. This corresponds
to the expected CO emission from Arp 220, or about 2$\times 10^{10}$
M$_\odot$ of H$_2$ using the standard Galactic conversion factor. 
The redshift ranges,  sensitivities, mass limits,
comoving volumes, 
and expected number of sources, are listed in Table 1. The expected
number of sources is derived using a comoving density of 
10$^{-3}$ Mpc$^{-3}$ for high z sources with 
350 GHz  flux densities $\ge$ 1 mJy, ie. sources with luminosities
comparable to, or greater than, that of  Arp 220 
(Smail et al. 1999, Carilli and Yun 1999b). 

Models of cosmic structure formation predict a  cosmic `dark age',
just prior to the first formation of stars in the universe,
sometime between $z$ = 6 and 15 (see Rees \& Meiksin, this volume). 
The New VLA will probe the limits of this dark age, 
revealing  radio continuum and molecular line emission 
from the first generation of star forming galaxies in the universe. 

\vskip 0.2in
\begin{center}
 Table 1: A search for redshifted CO emission using the New VLA
  at 26 to 34 GHz. 
  \begin{tabular}{|c|c|c|c|c|} \hline
Transition & Redshift &  Mass limit (5$\sigma$) 
& Comoving Volume  & Number of sources \\ 
~ & ~ &   $\times 10^{10}$ M$_\odot$ & Mpc$^{3}$ & ~ \\ 
\hline
CO(1--0) & 2.3 - 3.4 & 2.4  & 1300 & 1.3 \\
CO(2--1) & 5.8 - 7.8 & 2.4 & 1400 & 1.4 \\
CO(3--2) & 9.0 - 12 & 3.3 & 1450 &  1.4 \\ \hline
  \end{tabular}
\end{center}
\vskip 0.2in

\section*{Acknowledgements}
C.C. acknowledges support from the Alexander von Humboldt society.
The National Radio Astronomy Observatory is a facility of Associated
Universities, Inc., operated by the National Science Foundation.

\section*{References}

Baltz, E., Gnedin, N., \& Silk, J. 1998, ApJ (letters), 493, 1

Barger, A., Cowie, L., \& Sanders, D. 1999, ApJ (letters), 518, 5

Barger, A., Cowie, L., \& Richards, E. 1999, astroph. 9907022

Beck, R., Brandenburg, A., Moss, D. Shukirov, A., \& Sokoloff,
D. 1996, ARAA, 34, 155

Benford, D. J., Cox, P., Omont, A., Phillips, T. G.,
\&\ McMahon, R. G. 1999, preprint (astro-ph/9904277)

Blain, A.W. 1999, astroph 9905248

Blain, A.W., Kneib, J.P., Ivison, R.J., \& Smail, I. 1999,
ApJ (letters), 512, 87

Blain, A.W., Smail, I., Ivison, R., \& Kneib, J.P. 1999, MNRAS, 302,
632 

Blain, A.W., \& Longair, M.S. 1993, MNRAS, 264, 509

Calzetti, D. and Heckman, T. ApJ, in press, astroph 9811099

Carilli, C. L. \& Yun, 
M. S. 1999, ApJ, 513, L13

Carilli, C.L. \& Yun, M.S. 1999, ApJ,  submitted

Carilli, C.L., Menten, K.M., \& Yun, M.S. 1999, ApJ (letters), 521, 25

Cimatti, A., \&reani, P., Rottgering, H., \& Tilanus, R. 1998,
Nature, 392, 895

Condon, J.J. 1992, ARAA, 30, 575

Cowie, L., \& Barger, A. 1999, astroph. 9907043

Downes, D. et al. 1999, A\&A, submitted

Eales, S. et al. 1999, ApJ, 515, 518

Guilloteau, S., Omont,A., 
McMahon, R.G., Cox,P., \& Petitjean,P. 1997, A\&A, 328, L1

G{\"u}sten, R. et al. 1993, ApJ, 402, 537

Holland, W. et al. 1999, MNRAS, 303, 659

Kreysa, E. et al. 1999, SPIE, 3357, 319

Madau, P. et al. 1996, MNRAS, 283, 1388

Ohta, K. et al. 1996, Nature, 382, 426

Omont, A. et al. 1996a, A\&A, 315, 1 

Richards, E. 1999, ApJ, 513, 9

Saunders, R.S. et al. 1990, MNRAS, 242, 318

Smail, I., Ivison, R., Blain, A., Kneib, J.P., \& Owen, F. 1999,
astroph 9906196

Smail, I. et al. 1999, MNRAS, in press, astroph 9905246

Solomon, P.M., Radford, S.J.E., \&\  Downes, D. 1992,
Nature, 356, 318

Downes, D. \& Solomon, P. 1998, ApJ, 507, 615

Steidel, C., Adelberger, K., Giavalisco, M., Dickinson,
M., \& Pettinin, M. 1999, ApJ, 519, 1  

Tan, J.C., Silk, J., \&\ Blanard, C. 1999, ApJ, in press

Yun, M.S., Carilli, C.L., Kawabe, R., Tutui, Y., Kohno,
K. \&\ Ohta, K. 1999, ApJ, submitted

\end{document}